# Falling for Phishing: An Empirical Investigation into People's Email Response Behaviors

*Completed Research Paper*

**Asangi Jayatilaka**
CREST – the Centre for Research on
Engineering Software Technologies
School of Computer Science,
The University of Adelaide, Australia
asangi.jayatilaka@adelaide.edu.au

**Nalin Asanka Gamagedara Arachchilage**
School of Computer Science,
The University of Auckland,
New Zealand
nalin.arachchilage@auckland.ac.nz

**M. Ali Babar**
CREST – the Centre for Research on Engineering Software Technology
School of Computer Science,
The University of Adelaide, Australia
ali.babar@adelaide.edu.au

## Abstract

*Despite sophisticated phishing email detection systems, and training and awareness programs, humans continue to be tricked by phishing emails. In an attempt to better understand why phishing email attacks still work and how best to mitigate them, we have carried out an empirical study to investigate people's thought processes when reading their emails. We used a scenario-based role-play "think aloud" method and follow-up interviews to collect data from 19 participants. The experiment was conducted using a simulated web email client, and real phishing and legitimate emails adapted to the given scenario. The analysis of the collected data has enabled us to identify eleven factors that influence people's response decisions to both phishing and legitimate emails. Furthermore, based on the user study findings, we discuss novel insights into flaws in the general email decision-making behaviors that could make people susceptible to phishing attacks.*

**Keywords:** Phishing, Think-aloud study, Qualitative, Email response decisions

# Introduction

A successful phishing email attack can trick users into unintentionally disclosing their valuable information and install malware that can disturb the normal operations of a computer system (Arachchilage et al. 2016; Hanus et al. 2021). Organizations incur significant financial and reputation losses because of staff falling for phishing email attacks. The FBI's Internet Crime Report shows that in 2020, over $1.8 billion in losses was incurred due to business email compromise attacks alone—far more than for any other type of cybercrime (FBI 2020). Therefore, the importance of successfully detecting phishing emails has resulted in many phishing detection tools and techniques, which are continuously being improved (Gupta et al. 2017). However, even the most sophisticated anti-phishing tools and techniques can be inaccurate, where some phishing attacks are missed, and some genuine items are flagged as phishing attacks; therefore, they cannot

be considered a comprehensive solution to protect users from sophisticated phishing attacks (Gupta et al. 2018; Nicholson et al. 2017).

In terms of phishing emails, clicking on links, downloading attachments, or replying can be considered unsafe response decisions (Lawson et al. 2020; Parsons et al. 2013). Attackers often leverage human vulnerabilities to execute successful phishing email attacks. Furthermore, they are becoming more sophisticated by learning new techniques and changing their strategies (Arachchilage et al. 2016; Li et al. 2020). Therefore, there is increased attention on phishing training and awareness mechanisms (Jampen et al. 2020; Nguyen et al. 2021), including gamified approaches to educate users and enhance their capacity to thwart phishing attacks (Silic and Lowry 2020). Nevertheless, the percentage of successful attacks is still on the rise. Karjalainen et al. (2020) highlight the importance of understanding employees' current practices to explain why they may not comply with information systems security procedures. Similarly, in-depth insights into how people interact with emails is vital to unravel why they still fall for phishing email attacks.

Having a deeper understanding of user behaviors concerning email responses could better inform the design of any human-centric tool or technique (Albakry et al. 2020). However, there is a gap in the Information Systems (IS) literature with respect to the thought process a user goes through when deciding how to respond to their emails. A better understanding of the factors usually involved in making decisions about email communications can help understand why phishing emails still work and how best to mitigate the risks (Downs et al. 2006; Williams and Polage 2019). To address this research gap, we have formulated our research question as, **"What factors influence people's response decisions when reading their emails?"**.

We conducted an empirical investigation through a "think-aloud" role-play experiment and follow-up interviews to answer the above research question. While some of our findings confirm what was previously known, our study offers a significant amount of new and in-depth insights to advance the knowledge and understanding of human behavior that may lead to leverage against email-based phishing attacks. For example, our study reveals that there can be a disconnection between the people's judgment about the email's legitimacy and the email response; hence only focusing on people's email legitimacy judgments as undertaken in previous studies (Jones et al. 2019; Wang et al. 2016; Wen et al. 2019), is not adequate to obtain an holistic picture of why people fall for phishing emails. Succinctly, the contributions of this research are as follows:

- We identify eleven factors that influence people's email response decisions while reading their emails. These factors are: i) sender legitimacy; ii) perception about links in emails; iii) need for validation; iv) familiarity of the email title and body; v) professionalism of the email title and body; vi) emotional attachments with emails; vii) perceived likelihood of receiving an email; viii) length and granularity of information, ix) previous phishing experiences; x) sense of security from auxiliary security content; and xi) individual habits.
- Our findings provide novel insights into flaws in the general email decision-making behaviors that could make people susceptible to phishing attacks. These novel insights include: i) the possibility for a disconnection between a judgment on email legitimacy and the email response; ii) unsafe ways of validating email content; iii) role of emotions in email response decisions; iv) issues with identifying email link legitimacy; v) difficulties in deciding sender legitimacy; vi) issues in applying knowledge gained from formal education and lessons learned from exposure to phishing emails; and vii) effects of pre-conceived judgments in people's response decisions.

## Related work

This section highlights the literature related to users' susceptibility to phishing attacks and highlights the lack of literature that emphasizes users' thought processes related to reading emails.

Most of the studies that investigate understanding why people fall for phishing attacks focus on phishing websites or phishing URLs (Abbasi et al. 2021; Albakry et al. 2020; Alsharnouby et al. 2015; Iuga et al. 2016). Only a limited number of studies have been conducted in the phishing email context, where researchers have mostly looked into the demographic or personality characteristics of people who fall for phishing attacks (Aldawood and Skinner 2018; Hanus et al. 2021; Jones et al. 2019; Kleitman et al. 2018; Lawson et al. 2020; Li et al. 2020; Lin et al. 2019; Petelka et al. 2019; Sheng et al. 2010). For example, Lin et al. (2019) focused on susceptibility to spear-phishing emails and found that older women show the highest susceptibility for simulated phishing emails used in their experiment. Hanus et al. (2021) showed that various demographic characteristics such as age, salary, area of residence, nature of job-position, and computer access affect falling for email phishing attacks.

Several studies have investigated behavioral responses to phishing emails in order to further explain why people fall for them (Downs et al. 2006; Greene et al. 2018; Parsons et al. 2013; Williams and Polage 2019). Most research in this direction has investigated developing hypotheses based on existing theories borrowed from other fields and later testing their validity using user studies (Jaeger and Eckhardt 2021; Luo et al. 2013; Moody et al. 2017; Shahbaznezhad et al. 2020; Wang et al. 2016).

The work of Jaeger and Eckhardt (2021) is based on a theoretical model, developed using situation awareness research and protection motivation theory, that explains how security awareness is affected by individual and system-level factors and how situational information security awareness influences subsequent threat and coping appraisals. Wang et al. (2016) have developed and tested a model for over-confidence phishing email detection considering different aspects, including cognitive and motivational factors. Their findings draw attention to the issue of relying on judgmental confidence to drive user behavior. Shahbaznezhad et al. (2020), using diverse theories such as socio-technical theory and protection motivation theory, develop and test a framework that describes the factors that affect the intention toward non-compliance with e-mail security policy and phishing attacks. Moody et al. (2017) identify situational and personality factors that explain why certain people are more susceptible to phishing email attacks through a review of literature in the IS field, as well as the fields of psychology and communication. They then test those through empirical studies to explain the possibility for a person to fall prey to a phishing email attack. Their findings suggest that emails sent from a known sender significantly increase user susceptibility to phishing attacks. They also find that trust and distrust can be significant predictors of phishing email susceptibility. On the other hand, Greene et al. (2018) followed up with staff members within a government research institution to explore why they respond to emails in different ways during embedded phishing awareness training. Their findings highlight that clickers and non-clickers could interpret the same cues in different ways, depending on the alignment of the user's work context and the premise of the phishing email.

Unfortunately, the existing research has several limitations. Firstly, most of this research uses images of the emails in the experiment; however, this could detach participants from their naturalistic setting, which could affect their decision-making process. As also pointed out in previous work (Charters 2010; Downs et al. 2006; Kleitman et al. 2018; Parsons et al. 2013; Wang et al. 2016), when using images of emails, it is not feasible to understand whether or not people use the link URL information shown in the status bar since researchers cannot easily ask participants about it without alerting them to its relevance. Secondly, a common approach is to utilize surveys to obtain information from the users about their habits and traits and/or explanations for decisions that they have made (Greene et al. 2018; Moody et al. 2017; Wang et al. 2017). However, such self-reported surveys may not provide rich information about how users make decisions and may not be accurate, as users could be merely justifying their actions when answering the survey questions for decisions already made (Richardson 2004). Jaeger and Eckhardt (2021) have used an eye-tracking software in their experiment to reduce the limitations of using a survey. Thirdly, most research considers users' decisions about email legitimacy to be binary (e.g., phishing or legitimate) or user actions to be pre-defined. For instance, Parsons et al. (2013) allow users to select the action they would perform for a given email out of four pre-defined actions: a) leave the email in the inbox and flag for follow up; b) leave the email in the inbox; c) delete the email; or d) delete the email and block the sender. However, this approach may not align with people's natural decision-making processes. Fourthly, these studies have not explored the detailed thought processes users go through when making email decisions.

In summary, while the aforementioned research has provided much-needed insights into this field, further investigation into users' thought processes when responding to emails is imperative. The design of any anti-phishing tool/educational intervention will lack underpinning science without a detailed understanding of people's decision-making processes and associated flaws, which renders such tools and interventions less effective in making a notable difference in the way we combat phishing email attacks. To address the gap, our goal in this research is to identify what factors influence people's response decisions when reading emails. This will enable a better understanding of flaws in decision-making processes to answer why people fall for phishing emails.

# Methodology

We used a scenario-based, role-play think-aloud experiment and follow up interviews (Charters 2010) to collect data that could be analyzed to understand how people make response decisions while reading emails. The role-play enables researchers to study phishing without conducting an actual attack. The think-aloud method allows researchers to understand the participant's thought processes rather than only identifying their final actions. The follow-up interviews help to expand the think-aloud results and may allow the participants to "validate" the researchers' interpretation of their think-aloud statements (Charters 2010). Permission for the study was taken from the Human Research Ethics Committee of the University of Adelaide. The summary of the research methodology is illustrated in Figure 1.

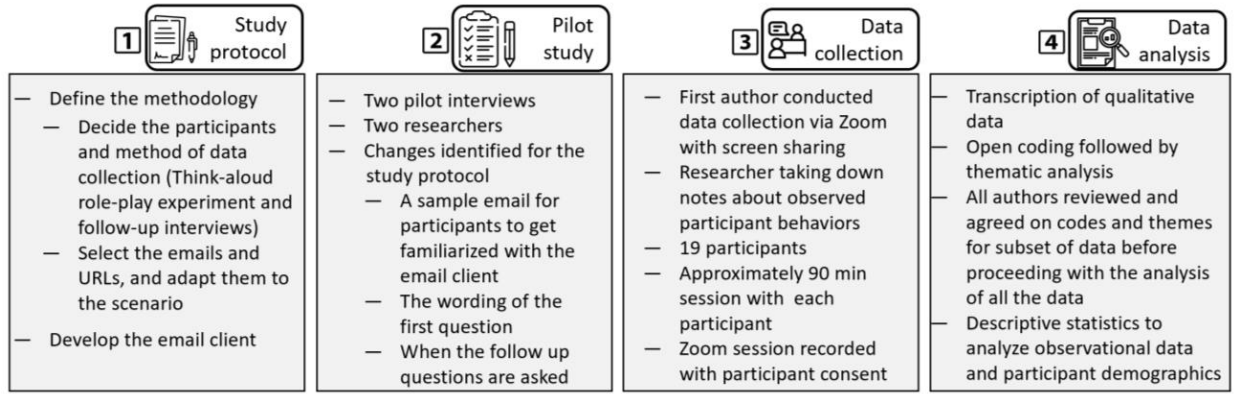

**Figure 1. Summary of the research methodology**

**Simulated email client:** We implemented a simulated web email client[1] for the study (Figure 2). The use of a simulated web email client allowed the participants to engage naturally with the task so they could easily contextualize their opinions when thinking-aloud. Furthermore, this set-up allowed the researchers to observe the participants' behaviors in the wild, in addition to what they explained. Previous studies have shown that most people who decide to click on a phishing link would already go on to disclose information to phishing websites (Sheng et al. 2010). As the focus of our work is to understand how people make email response decisions while reading their emails, the browser-based email client was prevented from navigating to any URL destination when a participant clicks on a link in the email body. The users were only shown a "link clicked" alert message when they clicked on any email link, regardless of whether the email is a phishing attack or not.

**Email selection, adaptation and assignment**: We composed 12 phishing and 12 corresponding legitimate emails by adapting real emails to a given scenario. These emails were selected from different life domains to increase diversity. Examples of the emails include Amazon order confirmations, Facebook alerts, University password expiry, PayPal notifications, mobile bills and shared Google sheets. The legitimate emails were mainly sourced from the researchers' personal email correspondence. The real phishing emails, dated from 2016 - 2020, were sourced from various venues including the UC Berkeley phishing archive[2] and Sensors tech forum[3].

A number of strategies that attackers use to create fraudulent emails were included in the selected list of emails (Misra et al. 2017). These include but are not limited to mimicking the sender's email address, providing a personal salutation, including the signature of the sender, emails creating fear and urgency, mimicking the appearance of a legitimate email, and URL obfuscation. Phishing URLs used in the study were adopted from real phishing links stated in PhishTank[4]. Recent research on URL obfuscation points to six ways attackers masquerade links to phishing sites in emails (Fernando and Arachchilage 2020; Garera et al. 2007). We included two phishing emails for all except one URL obfuscation technique, pointed out by Fernando and Arachchilage (2020). To materialize *obfuscating with HTTPS Schema*, the browser needs to navigate to the destination to show the encrypted icon on the address bar. However, as explained before, our email client prevented the user from navigating to external destinations, hence we omitted this obfuscation technique. We also included two emails that requested recipients download attachments and two emails that requested recipients reply back.

---

[1] https://github.com/asangi234/mymail

[2] https://security.berkeley.edu/education-awareness/phishing/phishing-examples-archive

[3] https://sensorstechforum.com/

[4] http://phishtank.org/

The email client was pre-populated with 12 emails. We assigned emails to the participants as follows. A specific phishing email and the corresponding legitimate email were considered as a pair. If a participant receives both a legitimate and a similar phishing email, they can easily compare and contrast the emails, which could lead to biased decisions. Considering a single email pair, we randomly assigned the legitimate email to 50% of the participants and the phishing email to the remaining 50%. This approach ensured that: i) a participant will not receive a legitimate email and its corresponding phishing email; and ii) given an email pair, half of the participants will receive legitimate, and the remaining half received the corresponding phishing email, which should lead to unbiased analysis.

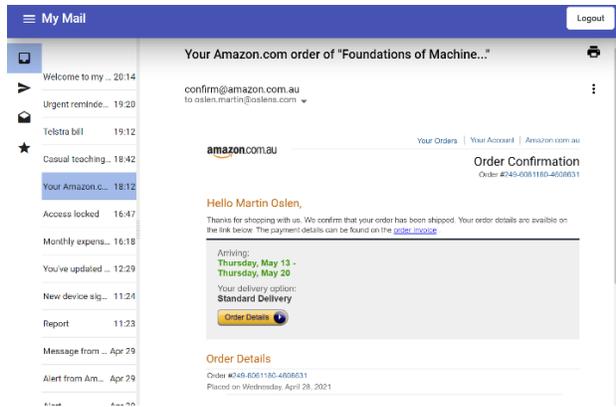

**Figure 2. Simulated web email client**

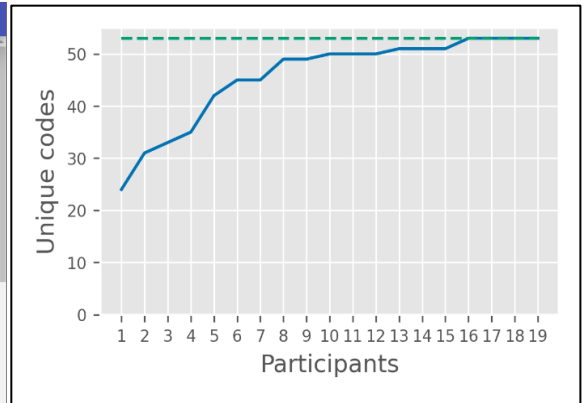

**Figure 3. The number of unique codes for each participant**

**Pilot study**: We conducted two pilot studies, which allowed us to identify three changes that were incorporated into the main study later. Firstly, during the first pilot, we identified that the initial question we asked (i.e., "how legitimate do they think a given email is?") restricted the participants from showing their natural behaviors. This question made the participants first decide on the email's legitimacy without much consideration and later provide justifications to support their decisions and then indicate an action that they would take that aligns with what was said before. Therefore, we decided to change the initial question to "how do you feel on receiving this email?". During the second pilot, we observed that this strategy allowed the participants to behave more naturally and talk openly about what comes to their mind when looking at the email, later explaining the actions they want to perform regarding the email. Secondly, the pilot interviews also made us understand that it would be good to provide a practice email for the participants to be comfortable with the set-up before commencing the think-aloud experiment. Thirdly, we made changes to the way the follow-up questions were asked. In the first pilot interview, we asked all the follow-up questions after finishing the think-aloud study. However, we identified that it is more effective to ask the follow-up questions after each email, as it is easier for the participants to answer the questions while their memory is still fresh.

**Data collection:** Think-aloud studies are often in-depth; hence typically involve a small number of participants (Charters 2010). For this study, by distributing flyers, we recruited 19 students at the University of Adelaide with various academic backgrounds. The participants were asked to fill out a questionnaire about their demographic information. Eight participants were male, and 11 were female. Twelve participants were undergraduate students (12), and the remaining (7) were post-graduate students. The participants were distributed across the Faculty of Arts (5), Faculty of Health and Medical Sciences (5), Faculty of the Professions (4) and Faculty of STEM (5). Most of the participants were 25 years of age and younger (14) and only 5 participants were over the age of 25. For the role-play, the participants were asked to assume they are a fictitious person named "Martin Oslen" and that they were viewing his email inbox. Background information on Martin was provided to the participants: they were asked to read and debrief the researcher. The participants were allowed to retain and refer to the document that describes the scenario throughout the study. Then the participants were provided with the URL to the email client with the user name and password. They were asked to share their screen with the researcher while going through the emails. The researcher provided basic instructions on the email client, and the participants were allowed to go through a sample email and interact with the email client. We explained that participants are free to perform any action (e.g., clicking on a link); however, this will only generate an alert message (e.g., link clicked). We did not explicitly instruct the participants on the techniques they can use to identify the URL's actual destination (e.g., hovering their mouse over a link to see the destination URL shown in the status bar), as that could influence their usual behavior.

The participants were asked to explain how they feel on receiving an email. They were asked to express their opinions in detail on the emails they were presented with. They were then asked to explain the actions that they took, with the reasons. We explained to the participants that the goal of the research is not to measure how well they perform but to understand how they make decisions. After each email, the participants took part in a short open-ended interview that allowed the researcher to ask follow-up questions. At no point did we restrict participants to deciding whether an email is phishing or not, as the goal was to provide the flexibility to them to make any decision they chose while reading the emails. The participants were given $20 voucher for participation.

**Data analysis**: The collected data was transcribed for thematic data analysis (Braun and Clarke 2012). We used the Nvivo™ data analysis software to organize, code and analyze the transcriptions. The thematic analysis framework identified by Braun and Clarke (2012), was used for the analysis. The first author conducted an initial open coding of the first six transcripts. The first author and other authors discussed those initial findings. As the study evolved, we proceeded with our analytical approach and revised already existing codes. The first author aggregated the codes into themes by following the guidelines gives by Braun and Clarke (2012). Then, themes and corresponding quotations for the first six sessions were reviewed by the second author, along with the interview content of the transcripts. Any disagreements were sorted out before moving further into in the analysis. We also had frequent iterative group meetings with all the authors throughout the process to ensure consistency and rigor in interpreting the data. We achieved saturation through this process (i.e., a state where little fresh information emerges from subsequent think-aloud sessions) because all the main themes had been uncovered across the 19 participants (see Figure 3).

# Study findings

Thematic analysis revealed eleven factors that explain what influences users' email response decisions when reading their emails. These are described below. Furthermore, based on the identified factors, we present how people can be susceptible to phishing attacks due to flaws in their decision-making processes. When quoting the participants, we indicate whether the quotation is in the context of a phishing (P) or a legitimate email (L), where relevant. In an attempt to provide further insights into the prominence of the identified factors in the underlying data, we report the percentage of the participants linked to each theme.

### *Sender legitimacy*

All participants (100%) consider the sender legitimacy when making email judgements. The participants tended to become suspicious when the sender domain is not relevant. However, we also identified inappropriate techniques employed by the participants to assess the sender legitimacy.

People often rely on the sender email address alone to make email decisions. The feeling of a legitimate sender makes them overlook the cues that suggest that an email is a phish. For instance, one participant explained that he would trust an email that appears to be coming from the University even if the email does not address him properly: *"Why is it saying that dear, like the entire email [as the salutation]? It would usually say, Dear Martin ... But I guess that it must be from the University, so seems like it's legitimate"* [P11—L]. The participants tend to trust emails that appear to be come from personally known people or their organization, and even tend to fully trust such emails without looking into the other parts of the email:

*"It depends on the sender. So, if my friend, family person sent an email. So that's not a problem for me ... University email is protected by the university"* [P01—L].

The participants fail to recognize and interpret different parts of the email address separately. They even make decisions only by looking into certain parts of a sender's email address (e.g., domain, sub-domain and user name). For example, one participant explained that having "org" in the sender's email address makes him feel safe: *"I just focus on this part like .org that makes me feel like probably like professional email address"* [P18—P]. Another participant explained that he trusts the sender's address because he thinks that it comes from the country he lives in, even though he knows that usually, that sort of email comes from PayPal.com: *"The sender email address is newsletter.com, I would trust that... I guess it sounds normal. It's because it's in [country he lives]... I guess normally it comes from Paypal.com"* [P15—P]. Some participants got confused with the subdomain and were reluctant to trust the email: *"What is this handle [sub domain] anyways? ... Does it have anything to do with the bank? I have no idea"* [P12—L]. The participants at times explained that having no-reply as the username of the email address made them feel safe for certain emails: *"I would check, and I feel like the no-reply makes me feel safe instantly"* [P16—L].

The participants at times made assumptions about the sender by only looking at the email name and/or email addresses specified in the email body: *"Okay so this is my partner [pointing to the email name] and I am pretty sure this is her email address [pointing to the email address in the email body]"* [P4—P]. We also observed that the participants could misinterpret the reply-to address. They make decisions about the sender legitimacy based on how they perceive the reply-to address: *"Oh I don't know what this is [reply-to address]. Because that's different from the emails on the role-play scenario sheet [Martin's wife email address]. But yeah, let's say it looks like a normal email"* [P9—P]. Sometimes they can get confused with the reply-to address, thinking it is the sender's email: *"This email is there [pointing to the reply-to address] Mary at dot org [Mary is Martin's wife and her email address is mary.oslen@oslens.com]"* [P1—P].

**Susceptibility to phishing emails based on this factor:** i) people can be deceived by spoofed email addresses with known senders or known domains; ii) people can be deceived by convincing email names, domains, sub-domains, user names related to the sender's email address and/or a reply-to addresses in phishing emails; and iii) people can be deceived by having a reliable-looking sender address in the email body to indicate that the email is coming from that sender.

### *Perception about links in emails*

Perception about email links influenced all participants' (100%) response decisions. The participants tend to believe that all phishing emails try to trick people to make them click on links. As a result, people can feel less vigilant when there are no links in the email: *"Without the links, then you don't have like those things that people are trying to get you to click on to, you know, download the viruses and what not"* [P09—L].

Where links are available, the participants tend to decide the URL legitimacy based on the link appearance. For example, one participant explained how he determines the link destination based on the text in the button: *"This button says it will take me to the website. I am able to review this unusual logging, I would believe that it would take me to a web site that I will review log in"* [P04—L]. Some participants tend to feel safe when they see the URL text in the email body instead of seeing a button: *"Usually it would say that www dot uber slash something change password something. That might be a bit more convincing than just clicking on the link that says change password now [button]"* [P11—P]. However, even when the URL is visible in the email body, the participants may not be able to recognize its legitimacy correctly. For example, one participant explained that he is skeptical about clicking the link due to the numbers at the end of the URL text: *"It is very common for an URL to have a long string of numbers. I saw this on several websites. It is this part the end makes me uncertain about this URL"* [P4—L]. This shows a lack of knowledge about URL structures. The network communication protocol is also observed to be leading to wrong judgements about URL legitimacy. Some participants explained that they feel safe when they see "HTTPS" as the network communication protocol compared with "HTTP" in the link text simply because they believe that HTTPS indicates a secure website: *"But what gives me a bit of confidence is this HTTPS secure server. So secured servers are unlikely, to be phishing sites so that would give me strength"* [P10—L].

The participants expressed that they feel safer clicking on specific links if the email does not mandate them to click. For example, one participant explained he feels safe about a mobile bill as the email is not asking him to click on any particular link: *"It was not asking for anything particularly. There was huge trust on me. It was not asking me to click a link … It's sort of email that you are welcome to ignore"* [P19—P].

Only seven participants considered the link destination at least once to decide the link reliability. They either hovered over the link or copy-pasted the URL, but none of them consistently looked into the link destination for all their emails. Furthermore, even after identifying the link's actual destination, sometimes the participants made wrong decisions about the URL legitimacy. We observed that most of these wrong decisions could be attributed to the lack of knowledge about URL structure and URL obfuscation. For example, sometimes participants are content when they see few relevant keywords in the destination URL:*"Yeah, it's popping up on the bottom left-hand side. It says sign in dot amazon dot com dot au slash and a few more things. But um, yeah, just that sign in dot Amazon, which I guess makes sense"* [P9—P].

**Susceptibility to phishing emails based on this factor**: i) people can be deceived by phishing emails composed without any links but requesting reply or download attachments; ii) people can be deceived by phishing emails with legitimate looking buttons or URLs in the email body; iii) people could make wrong judgments about URL safety by looking only at the network communication protocol mentioned in the URL; iv) people can be deceived by phishing links that appear as non-mandatory; v) people can be deceived by phishing emails having a destination URL that is different from the URL text; vi) even people who understand techniques to identify the URL destinations can still be susceptible to manipulation as they fail to consistently apply those techniques in practice; and vii) even after looking into the destination URL, people can be susceptible to phishing due to a lack of awareness about URL obfuscation and URL structures.

## *Need for validation*

All participants (100%) expressed a need for validating certain emails before deciding the final action particularly when they were suspicious about the email or when they were extra vigilant even when they thought the email looked legitimate. While some techniques expressed by the participants to validate the email can be considered safe, some techniques, unfortunately, are unsafe.

The participants validate emails using information gathered externally to the email. This includes but not limited to using the relevant mobile application and visiting the website separately: *"I would go and check my online account, mobile app or through website and verify if this is the case"* [P10]. Both these techniques can be considered safe. However, participants also explained that they would search for information mentioned in the email on the internet before deciding the final action. For example, finding the logo, company information specified in the email on the internet could make participants believe that they have received a legitimate email. However, this is an unsafe way of validating the email content as attackers would also have access to that public information when creating emails.

Some people validate an email through information obtained from the email, though it is an unsafe approach. People assume that it is safe to click on links if they are not providing information on the landing page: *"I would click on 'View details' straightaway and if it is amazon, it will take them to the logging site no doubt. If it's not Amazon, that's a question"* [P10—L]. Some even want to reply to the email to verify: *"I probably would try to email back and see what I get as a reply"* [P12—P]. Some explained that they would not click the main link but would click the other links in the email to make a decision about the email legitimacy of the main link: *"What I would do if I received this email like it's kind of funny but I would first click this let us know right away [link] so it won't first take you to like deactivate email … If it says PayPal and then submit and all then I'll be sure that this is real and then I'll come back and then only I'll click on this link"* [P1—P]. Some participants indicated that they would click the links on the email to get help and more information about the email: *"But then at the bottom, it says, to get in touch, go to the PayPal website and click help, and that's probably what I would do"* [P17—P]. One participant explained that he would call the number given in the email and decide what he wants to do for the email depending on what he hears during the phone call: *"When I call this number, there should be a guided phone call you know-welcome press one press two press three. I mean in the past, I experienced a scam phone call as well. The volumes the quality of the phone call is not that great … If they are guiding you through all of these details and all that then I believe that it's it's more legit"* [P6—L].

**Susceptibility to phishing emails based on this factor**: i) people can be deceived through publicly available information that is mentioned in emails; ii) people can be deceived by legitimate-looking landing pages for phishing links. Although people may not provide their details on these landing pages, it could still download malware (i.e., malicious IT applications) to the victim's computer system. One form of malware is ransomware which encrypts the data on the victim's computer system such as files, folders, images etc.; iii) people can feel safe to click on secondary links available in the phishing emails to validate the main link without understanding that the secondary links could be phishing links as well; and iv) people can be deceived by seeing various ways (e.g. contact us links) to get help about the email in the email content.

### *Familiarity of the email title and body*

All participants (100%) tend to believe and act upon emails that are familiar to them. Some of the aspects they mentioned include familiar layouts, logos, title, writing styles and footer. For example, a participant explained she would not second guess any University email that has a familiar look and feel: *"I mean for all the university emails, that have the logo and they have this pattern of putting this with university things … I'm confident"* [P7—L]. The participants overlooked cues that suggest an email is phishing when its look and feel is familiar to them. This strategy sometimes made the participants suspicious about legitimate emails, especially when something unfamiliar is present in an email. For example, a participant explained she suspects a legitimate email as being phishing because of the unfamiliar logo: *"This is definitely not the logo for the bank. So even though everything else seems kind of real, kind of legit, because of the logo itself I probably will not trust this that much"* [P12—L].

At times, when participants are not familiar with the email, they tend to look up elements on the internet to find out information relevant to the email before taking any decisions. For example, one participant explained that she is not familiar with the logo given in the email, she would search it up on the internet. After searching the internet, she decided to trust the email as she found a logo on the internet that is like the one available in the email: *"Okay, well I mean I can't remember what the color of the bank logo is. I am just going to search it up. I am guessing this is the exact same one here as the one here"* [P18—L].

**Susceptibility to phishing emails based on this factor**: i) people can fall for emails that have a similar look and feel to the emails they have seen before; ii) people who are not familiar with an email could be still susceptible to manipulation if they match the email content with what is available on the internet.

### *Professionalism of the email title and body*

All participants (100%) expressed that they look for the professionalism of different aspects of an email, including the professional layout, logo, other organizational related links, footer, button appearance and spelling correctness. For example, one participant decided to ignore the email after observing spelling errors: *"I think this is a little bit fake because this is `comrn', they have very smartly made `rn' seem like `m'… So I think this is fake"* [P3—P]. Another participant tends to trust a phishing email due to the professional look and feel of the email created through other organizational related links available in the email: *"They have these links, contact links and all which would certainly be on a reliable source … So, these links make me feel comfortable"* [P1—P].

A recipient, at times, can overlook the cues that they notice in emails that suggest they are phishing emails because of the professional look and feel of those emails. For example, one participant explained that although not having the product information on the Amazon receipt confuses him, he would still trust and click the email link due to the email's professional look and feel. He further explained that he assumes that it would be a minor issue in the Amazon system that has resulted in the product details not being printed on the receipt: *"This looks like a very professional email … I am very struck on the fact that is paranoid of me that it has not got the product I bought"* [P4—P].

**Susceptibility to phishing emails based on this factor:** people could be susceptible to phishing emails that include a professional-looking title and the body.

### *Emotional attachments with emails*

It was clear that emotions play an essential role in how most participants (95%) respond to emails. The participants tend to immediately respond to emails due to emotional attachments, even without considering email legitimacy. For example, the participants express their enthusiasm to receive a job offer and desire to take immediate action without much consideration about the legitimacy of the email: *"And then probably especially because of my excitement and everything I wouldn't second guess it, I would just read the* *agreement" [P12—L].* Furthermore, curiosity could drive email response decisions. For example, the participants explained that they want to click on the given links due to their curiosity to learn more about the email: *"Because it says here, click here to stop this action now, I just want to know what's happening when I click" [P06—P].* The participants prioritize academic work over the other tasks. They may feel compelled to respond to emails that appear to from a supervisor to maintain good relationships even when they observe phishing cues: *"I'm not satisfied sending this email, but since I need to keep my degree like, hold my degree because he's the person in charge of my degree so, I would email [him]"* [P1—P]. Email content related to sensitive information and finances influences response decisions. For example, P11 explained that he was compelled to follow the link in the email, although it looked suspicious: *"I would have my bank account linked and everything … Even if you think that it's spam and everything, you still kind of likely to click on the link, although the link looks quite weird"* [P11—P]. At times, the email recipients can be annoyed when they identify a phishing email driving them to delete the email or mark it as spam. However, P06 said he can be so annoyed with the email that he would want to unsubscribe from such an email by clicking on the given link: *"I would want to click here because, I know that this is a scam email. But I want to unsubscribe from this newsletter"* [P6—P].

**Susceptibility to phishing emails based on this factor**: i) people can be susceptible to manipulation by being presented with phishing emails that elicit different emotions such as happiness, anxiety, work-related priorities, and curiosity; ii) people may not even think about email legitimacy before responding to email due to their emotions; iii) emotional attachments with the email could make people respond to emails even ignoring the phishing cues that they observe; and iv) emotions could drive people to click on links in emails even after identifying them as phishing emails.

### *Perceived likelihood of receiving an email*

The participants' (94%) response decisions could depend on how they perceive the likelihood of receiving the email. For example, if they had had a user interaction with the entity recently, then they are more willing to accept and act on emails coming from that source: *"Since I have been into this website and actually changed my password, I'll be expecting such email"* [P04—L]. At times participants falsely believed phishing emails are legitimate, overlooking the cues that suggested otherwise, based only on the perceived likelihood of receiving those emails: *"The timing that I got [the email], if I didn't do anything beforehand and it just came without me expecting it. That would be the first factor"* [P05—P].

**Susceptibility to phishing emails based on this factor:** i) people can be easily deceived by targeted emails (e.g., spear phishing) created by attackers who have knowledge about the user's context and interactions; and ii) people can be deceived by phishing emails that coincidently overlap with their expectations (e.g., periodic utility payments).

### *Length and granularity of information*

The participants (84%) tend to look for the length and granularity of an email before responding to them. When an email is long, the participants tend to think it is more legitimate: *"This is usually pretty long like the phishing email would usually be like very short and they will not, like, completely replicate what how a company send their email"* [P11—P]. The participants explained that they believe attackers will not try to compose lengthy emails: *"In my point of thinking like I think a spammer doesn't want to type this much to make a person fall into trap"* [P01—P].

When an email has detailed information of a situation, participants tended to have more trust in those emails: *"That'll be a legitimate email which describes what the problem is and what has happened, what actions are going to be taken and how you should contact them. It's not like we are launching the missile and click here to stop"* [P10—P]. Specifically, having customized details about the recipient in an email makes them trust the email. The participants believe that attackers will not be able to know the transactions or orders performed by them: *"I believe that Amazon.com they should protect my privacy information. So, then no one else would know that I have already ordered"* [P06—L]. One participant explained how the granular details given about an Amazon order convince her that it is a legitimate email: *"All these nitty, gritty details about; say the amount of tax you have to pay… I would trust it a bit more"* [P12—P]. Another participant explained that she thinks that attackers cannot access the names of users easily: *"They probably wouldn't have time to go target me specifically and send me an email with my name on it"* [P12—L].

**Susceptibility to phishing emails based on this factor**: i) people can be deceived by lengthy phishing emails; and ii) people can be deceived by targeted phishing emails (e.g., when a purchase order for a book is raised, the cyber attackers can specifically target the victim using that information) with detailed information related to the recipient.

### *Previous phishing experiences*

It was evident that participants (80%) previous exposure to phishing emails and formal phishing education affect their response decisions. People learn accurate and inaccurate strategies for detecting phishing emails by being exposed to phishing emails. For example, one participant explained how her past experiences with a phishing email made her think that phishing emails/URLs are composed of only numbers: *"From that specific email, I learnt that fake emails use computer generated number language …. I feel like if it was a legitimate email you would not have the URL with bunch of numbers. Real emails would have indication using language"* [P04]. We also noticed that people could get too scared of clicking links in any email due to past exposure to phishing emails: *"I won't click any links in my personal opinion for emails … My email was completely hacked. So, I had to remove that account, so I won't be clicking any links"* [P01—L].

Few participants explained how their phishing related formal education affected their response decisions: *"I also checked for the punctuation all because I've learned from my computer classes that spam emails have misspellings and punctuation"* [P4—L]. However, our results provide evidence that people do not always consistently apply their knowledge in practice. For example, P9 explained that she had been taught to hover over the link to see the actual link destination: *"Well, what I've been taught before is that if you hover over the link, like not clicking it, but you hover over it, it should tell you where it'll actually take you"* [P9]. However, through the interactions with the web email client, we observed that P9 hovered over the links only 60% of the time. Another participant explained that he had learnt that phishing emails are created to look very similar to legitimate emails and even the sender email can be spoofed: *"I've been told that [in phishing emails] the layout the fonts and everything exactly hundred percent same [as a legitimate email]. They [attackers] even sometimes change, this sender email to the same that the legit sender"* [P06]. However, throughout the role-play experiment, we noticed situations where this participant did not apply this knowledge in practice.

**Susceptibility to phishing emails based on this factor:** i) previous exposure to phishing emails could make people learn inaccurate strategies for identifying phishing emails; and ii) even though the participants may have the knowledge to identify phishing emails through formal education, they can be still susceptible to phishing because of not consistently applying the knowledge gained in practice.

### *Sense of security from auxiliary security content*

Sometimes participants (68%) tend to feel an email is safe when security-related information is provided in the email footer or body. For example, the participants tended to associate security-related information provided in the footer with the sender's intention to make the receiver's security a priority: *"It [email footer] just makes me feel a little bit more secure. The company cares about my privacy."* [P3—P]. Another participant explained that she thinks having a privacy statement in the email footer provides a sense of safety and allows her to place more trust in the email: *"Privacy statement here and this closing paragraph here already says strictly confidential. It makes me feel it is a bit more safe"* [P18—P].

The participants also tend to develop a sense of security when there is an indication that the email has been scanned by external scanning tools, even when they do not know much about those tools. *"This is sort of third-party email scanning anti-virus, Forcepoint. I don't know what this Forcepoint [is] but this sort of making me trust this is an email that has been scanned as no virus on this email"* [P6—P]. Another participant said she would be confident in clicking the link if she could find such a tool on the internet: *"I'm not sure of this text … But I'd go to Google and then check the Forcepoint website. If there's an actual website … then I can also be satisfied that this part of the message is true and full email is legitimate"* [P1—L].

**Susceptibility to phishing emails based on this factor**: people can develop a false sense of security through privacy statements, confidentiality statements and information about scanning tools available in phishing emails.

### *Individual habits*

At times (68%), the personal habits of the user come into play when deciding how they respond to their emails. Some participants would habitually click on any email link straightway without thinking: *"It's in general when someone shares a document to me, I can click it and look"* [P6]. For some participants, how they read and respond to phishing emails would depend on the day and time: *"It also depends on the time of day that I'm reading the email.. If I'm just chilling out just relaxing, and I will go through my emails leisurely"* [P5]. Some were extra vigilant and always and try to avoid clicking on links on any email: *"I wouldn't anyway click ... Unless I am very much sure, I wouldn't click or I wouldn't respond to any of the emails"* [P02]. On the other hand, some participants can be extra vigilant for certain types of emails. They explained that they are more cautious with alert emails and emails related to banking and money but appear not to be very cautious about emails linked to their family, work or education: *"Normally for this type of information ... I would deal with it on Facebook and not on email"* [P11—L].

**Susceptibility to phishing emails based on this factor**: i) people could respond to emails even without thinking about their legitimacy due to their habits; ii) people could be susceptible to phishing emails based on the time of day they look into their emails;and iii) people could be susceptible to phishing due to their pre-judgments about certain types of phishing emails.

## Discussion

Our study offers significant contributions to the field by providing in-depth insights into how people make email response decisions and how they are susceptible to manipulation due to flaws in the decision-making process. Like previous studies (Downs et al. 2006; Parsons et al. 2013; Williams and Polage 2019), our study highlights that people tend to trust emails targeted at them with detailed information and/or that look professional. Furthermore, our findings confirm previous studies' findings that suggest that the phishing email's alignment to user context could make people have more trust in those emails (Greene et al. 2018; Jaeger and Eckhardt 2021; Williams et al. 2018). The role of personal characteristics and habits in email decision-making is also highlighted on several occasions (Ayaburi and Andoh-Baidoo 2019; Jones et al. 2019; Moody et al. 2017; Vishwanath et al. 2011). While some of the current study findings confirm what was previously known, there are a significant amount of new or detailed insights into what can be considered missing pieces of the puzzle to understand why phishing email attacks still work. These are described below.

### *The possibility for a disconnection between a judgment on email legitimacy and the email response*

Our study revealed that simply understanding how people judge email legitimacy is not adequate to obtain an holistic picture of why people fall for phishing emails. Previous user studies on phishing susceptibility often focus on users' email legitimacy decisions (Jones et al. 2019; Wang et al. 2016; Wen et al. 2019). These studies assume people decide the email's legitimacy (i.e., phishing or legitimate) before responding to emails, and correctly identifying email legitimacy will make users less susceptible to phishing. While our study confirms this decision-making process, it provided further insights into people's email legitimacy decisions. There could be situations where there is a disconnection between the email response and the judgment on email legitimacy. Our results show that people could click on links, reply, and download attachments without even thinking about the email's legitimacy due to emotions and individual habits. Furthermore, our findings reveal that people could click on email links or reply to emails even after identifying phishing emails or when they are unsure of the email's legitimacy.

This finding could have significant implications for training and assisting users not to fall for phishing emails. The users have to be taught and assisted not only to identify email legitimacy before responding to emails but also to make safe response decisions.

### *Unsafe ways of validating email content*

Often previous research related to email phishing requests participants to decide whether a given email is legitimate or phishing (Jones et al. 2019; Wang et al. 2016; Wen et al. 2019). However, as described through our findings, people may not be willing to make a final decision on the email legitimacy while going through an email and may intend to validate the email before making the final decision. This has not yet been shown in the IS literature. Although validating an email can be considered a precautionary measure, our findings reveal that some techniques used to validate the email could be unsafe (e.g., calling the phone numbers given in the email to verify the content, clicking on the link to see the landing page and searching the internet to see if they find information similar to what is specified in the email) and result in people becoming victims of phishing attacks.

This finding has consequences as to how training is conducted in organizational settings, specifically validating the contents in an email. Furthermore, attention should be paid to what reliable tools can be provided to the users that can be used for this purpose.

### *Role of emotions in email response decisions*

Our findings reveal how people can be susceptible to manipulation by being presented with phishing emails that elicit different emotions such as happiness, anxiety, curiosity, and emotions arising from work-related priorities. A wealth of research has documented how specific emotions can affect individuals' perceptions, judgments, and behavior (Gulenko 2014; So et al. 2015). However, not much is yet known in the IS literature about how people's emotional attachments with emails affect their decision-making process and email response decisions. Our study provides insights into how the emotional attachments people develop with emails could influence their email response behaviors. Emotional attachments people develop with the received emails could drive them to take impulsive responses by not only disregarding the email legitimacy but also ignoring the phishing cues they notice in those emails. Furthermore, negative emotions (e.g., anger) after identifying a phishing email could make people perform unsafe actions (e.g., clicking on unsubscribe links) on emails without considering the possible consequences.

We anticipate that these findings can be used to design better anti-phishing education techniques and tools. Anti-phishing education and training should emphasize diverse and complex emotions that could arise from reading emails (e.g., emotions arising from work-related priorities, anger arising from receiving a phishing email) and how those emotions could alter human behaviours. Furthermore, identifying email emotions should be introduced as a step before acting on an email. Tools to combat phishing could utilize the advancements of sentiment and emotion recognition techniques to automatically assess the emotional aspects of emails, which could be used to guide users to take actions safely.

### *Issues into identifying email link legitimacy*

Our results provide novel insights into how people's perceptions about email links affect their response decisions. It is observed that the participants tend to have more trust in emails without links. For emails with links, they use unsafe techniques such as clicking on links that appear to be non-mandatory and consider the button or URL appearance to assess the link legitimacy. Furthermore, often, participants do not know how to identify the URL destination (e.g., by hovering over the anchor text and looking into the URL provided in the status bar), and those who know how to identify the URL destination but do not consistently apply their knowledge. As discussed in the Related Work section, this information was not available for previous studies. Our study findings further revealed that even after identifying the destination link, people may struggle to identify its legitimacy due to a lack of knowledge of URL obfuscation (Fernando and Arachchilage 2020) and URL structures. A recent study (Albakry et al. 2020) explored users' URL reading outside the email context also explains a similar finding regarding URL structures. Most people failed to interpret the URL structure correctly and thought that the URL would lead to the website of the organization whose name appeared in the URL, regardless of its position.

This finding has consequences for training users to identify phishing links embedded in phishing emails, especially with complex URL obfuscating techniques. Furthermore, the designs of email phishing warning messages should consider the possible users' lack of awareness about URL structures and URL obfuscation.

### *Difficulties in deciding sender legitimacy*

People often look at the header information to make email decisions (Luo et al. 2013; Moody et al. 2017; Pfeffel et al. 2019). While our results also confirm that sender legitimacy is one of the key factors in the email decision-making process, the study findings provide further insights into people's misconceptions when making judgments about the sender's reliability. Although people know that the email sender plays a crucial role in identifying email legitimacy, our findings demonstrate that they seem to lack knowledge as to how to use the email header information to identify any discrepancies correctly. For example, people often get confused with the structure of the sender's email address (i.e., domain, sub-domain, username), email display name, reply-to address, and email addresses specified in the email body, subsequently leading to flawed decisions about the sender's legitimacy.

The findings provide evidence that simply highlighting a sender's information, as done in the previous work (Nicholson, Coventry, and Briggs 2017), to make users aware of sender discrepancies may not be useful. Users should be made aware of how to interpret different parts of the email header and where/where not to focus to identify sender legitimacy.

### *Issues in applying knowledge gained from formal education and lessons learned from exposure to phishing emails*

Previous research highlights that formal education and previous phishing-related experiences could provide necessary skills in identifying phishing attacks (Arachchilage et al. 2016; Chen et al. 2020; Jaeger and Eckhardt 2021; Wright and Marett 2010). Our study findings support this claim and provide further insights into this aspect. Our findings suggest that previous exposure to phishing emails could make people learn both accurate and inaccurate strategies to identify phishing emails. Furthermore, exposure to phishing emails in the past could make them scared to respond to emails in any way, even when they feel emails are legitimate. Furthermore, our results provide evidence that even though formal education can make people acquire knowledge to identify phishing emails, they can still be susceptible to manipulation because of inconsistencies in applying the knowledge in practice.

These findings suggest that there should be ways (e.g., gamified approaches) to assist users in validating what they learn from being exposed to phishing emails is accurate. Furthermore, given the issues with applying knowledge gained from formal education into practice, we believe there should be more focus on designing technology-mediated nudging mechanisms (Petelka et al. 2019; Volkamer et al. 2017) so that users can be assisted to make better email response choices in the wild.

### *Effects of pre-conceived judgments in people's response decisions*

Previous research suggests that urgency arousing cues embedded in a phishing email are positively related to phishing susceptibility (Ayaburi and Andoh-Baidoo 2019; Williams et al. 2018). However, our findings point out that this may not always be true as people can be extra vigilant in some of these situations due to pre-conceived judgments about specific types of emails. For example, we observed how people could refuse to click on links in emails with urgency arousing cues due to pre-conceived judgments about social media alerts and alerts messages related to their banking or money, even when they feel the email is legitimate. However, they appear not to be very cautious with emails linked to their family, work, or education due to having a pre-conceived misconception that attackers will not launch their attacks through those types of emails.

This finding highlights the need to address users' misconceptions about phishing emails, make them aware of different forms of phishing email attacks (e.g., spear phishing (Hanus et al. 2021)), and continuously update them with new information.

## Limitations, Conclusion and Future Work

We have conducted an empirical study to gather evidence on what factors influence people's response decisions when reading their emails. The study findings open up the black box of the end-user decision-making process when deciding email responses. Based on the user study findings, we discuss how people are susceptible to manipulation, even in our controlled experiment environment, due to certain assumptions underpinning their decision-making approaches.

There are three main limitations in the current study. Firstly, the study participants consisted of University students, who do not represent the full range of email users. Previous research highlights that educational background in University students could reduce their susceptibility to phishing attacks (Gavett et al. 2017; Sheng et al. 2010; Williams and Polage 2019). This could be due to the factors associated with greater critical thinking skills and learning (Williams and Polage 2019). Therefore, it could be expected that other population samples will be more or differently susceptible to phishing attacks; however, future research is needed to confirm this. Secondly, similar to previous studies (Alsharnouby et al. 2015; Nicholson et al. 2017; Petelka et al. 2019; Wang et al. 2016), our participants were told from the beginning that they were taking part in a phishing related experiment. Although we explained clearly to the participants that the research goal is not to measure how well they perform but to understand how they make decisions, we want to highlight that the instructions may have primed the participants to examine emails more closely than perhaps they would do otherwise. Therefore, it can be expected that our participants behaved much more vigilantly than they would normally behave "in the wild". Apart from fulfilling the ethical and responsible research of informed consent, we argue that there can be several advantages in participants being aware that they are taking part in a phishing-related experiment focusing on their decision-making processors. For example, we observed how participants opened up to us easily about their previous exposures to phishing emails and the previous phishing education and how that affects their email decisions while reading their emails. Furthermore, informing the participants that they were taking part in a phishing-related experiment avoided any confusion that could be caused if the participants realize the nature of the experiment during the study, especially when the follow-up questions were asked to understand their behaviors in-depth. Thirdly, we quantified the themes only to provide some context to the reader about their prominence in the underlying data; however, without a large-scale study, we cannot comment on the generalizability of the provided statistics.

We believe that the current study's findings have the potential for theory building in phishing email detection in the future. Large-scale studies should be conducted to validate the current study's findings and investigate their generalizability. Furthermore, future research should look more closely into designing and evaluating effective anti-phishing education and training interventions (Silic and Lowry 2020; Wen et al. 2019) and nudging mechanisms (Nicholson et al. 2017; Petelka et al. 2019) by considering the in-depth insights into how users make email response decisions and flaws in those decision-making processes, as revealed through this study.